\documentclass[epj]{svjour}
\usepackage{graphics}
\begin{document}
\title{Searches for BSM (non-SUSY) physics at the Tevatron}
\author{Heather K Gerberich\inst{1} \ (for the CDF and D\O \ Collaborations) 
}                     
%
%
\institute{University of Illinois Urbana-Champaign }
\date{Received: date / Revised version: date}
%
%
\abstract{Results of searches at the Tevatron for physics (non-SUSY
and non-Higgs) beyond the Standard Model using $200\ $pb$^{-1}$
to  $480\ $pb$^{-1}$ of data are discussed. 
Searches at D\O \ and CDF for $Z'$, Lepton-Quark compositeness, Randall-Sundrum 
Gravitons,  Large Extra
Dimensions, $W'$, Leptoquarks and Excited Electrons
are presented here.
\PACS{
      {PACS-key}{discribing text of that key}   \and
      {PACS-key}{discribing text of that key}
     } 
} 
\maketitle
\section{Introduction}
\label{intro}

The discovery of anomalous behavior in data collected
at high energy physics experiments
could provide non-SUSY and non-Higgs explanations to
questions associated with the Standard Model and provide
deeper understanding to the fundamental particles and 
interactions in nature.  Such questions include 
whether quarks and leptons are composite particles, the
existence of extra dimensions, and the answer to
the hierarchy problem in the Standard Model (SM).

Generally, a search is approached by first understanding
the SM prediction for a given signal and detector backgrounds
which could mimic that signal.  Analyses are optimized
for signal, not according to model, prior to looking in 
the signal region of the data.  If no anomalous behavior
is found, the signal acceptances of various models can
be used to set limits.


\section{High Mass Dilepton Searches}
\label{sec:dilepton_diphoton}

High mass dilepton searches are experimentally
motivated by the small source of background, 
with the exception of the well-understood,
irreducible Standard Model
$Z/\gamma^*$ production.  Search results
can be used to study many theories:  extended
gauge theories ($Z'$), technicolor, 
lepton-quark compositeness,  large extra
dimensions (LED), and Randall-Sundrum gravitons.

\subsection{$Z'$}
\label{sec:Zprime}

The majority of extensions to the SM predict new
gauge interactions, many of which naturally result in the prediction
of neutral or singly charged bosons, such as a highly
massive ``$Z'$'' particle. 

\subsubsection{$Z'$ Searches using $M_{ee}$ and $\cos \theta^*$}
\label{sec:ZprimeMeeCosTheta}

\begin{figure}
\resizebox{0.5\textwidth}{!}{%
  \includegraphics{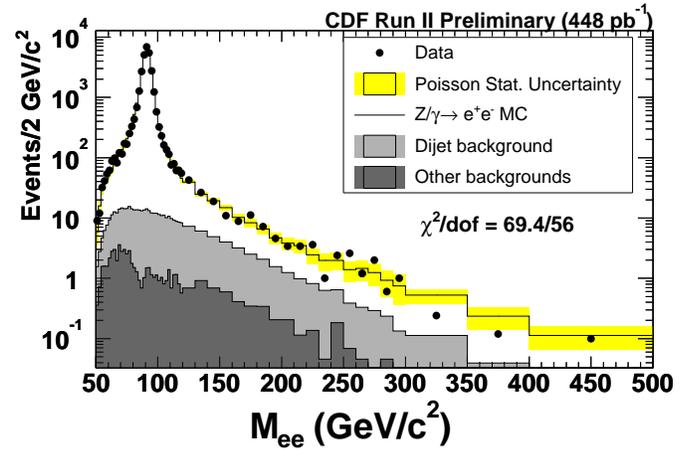}
}
\caption{Expected and observed dielectron mass distributions.}
\label{fig:CdfMee448}       
\end{figure}

\begin{figure}
\resizebox{0.5\textwidth}{!}{%
  \includegraphics{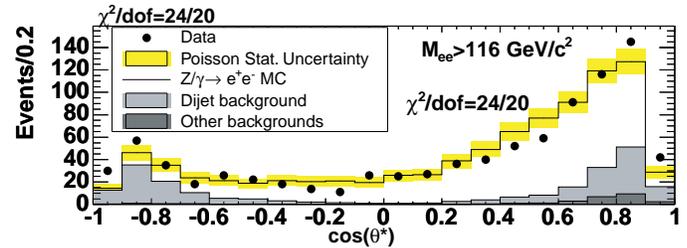}
}
\caption{Expected and observed $\cos \theta^*$ distribution for $M_{ee} > 116$ GeV$/$c$^2$.\label{fig:CDFCosTheta448}}
\end{figure}

Using $448\ $pb$^{-1}$ of data, CDF searched for $Z'$ production by studying
the distributions dielectron mass at high mass and the
angular distribution $\cos \theta^*$.
Figures~\ref{fig:CdfMee448} and~\ref{fig:CDFCosTheta448} show the
$M_{ee}$ and $\cos \theta^*$ distributions, respectively.

Having observed no evidence of a signal, limits at the 95\% 
confidence level (C.L.) are set for the sequential $Z'$\cite{bib:SeqZprimeModel}
and E6 $Z'$ models\cite{bib:E6ZprimeModel}, as shown 
in Table~\ref{tab:ZprimeTraditionLimits}. 
With $448\ $pb$^{-1}$,
using the $\cos \theta^*$ information effectively increases
the amount of data by $\approx 25\%$ for the sequential
$Z'$ model.

\begin{figure}
\resizebox{0.45\textwidth}{!}{%
  \includegraphics{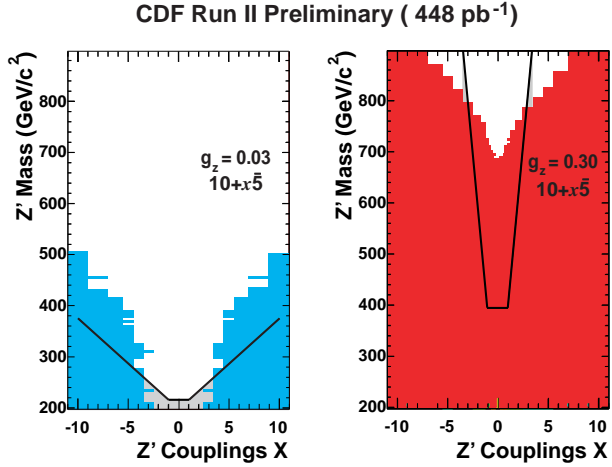}
}
\caption{Exclusion regions using a generalized formalism for $Z'$
searches.}
\label{fig:CarenaZprime}
\end{figure}

Additionally, a general formalism for $Z'$ which uses
$M_{ee}$ and $\cos \theta^*$\cite{bib:carena} and allows for new models
to be easily checked is studied.  The formalism
consists of four general model classes and are each defined by
three parameters: mass ($M_{Z'}$), strength ($g_{Z'}$) and
coupling parameter ($x$).  Figure~\ref{fig:CarenaZprime}
shows the CDF exclusion regions for one of the model classes for 
two values of $g_{Z'}$.  The area below the black curves represent LEP II
\cite{bib:carena} exclusion regions obtained via 
indirect searches for contact interactions.

\subsubsection{Traditional $Z'$ Searches}

CDF and D\O \ both performed ``traditional'' $Z'$ searches
which focus on the dilepton mass distributions.  All three
channels - electron, muon, and tau - were studied with no 
evidence for a signal beyond the Standard Model expectations.
Table~\ref{tab:ZprimeTraditionLimits} shows a summary of
the limits set at the 95\% C.L. for various $Z'$ models.

\begin{table}
\caption{Limits from CDF and D\O \ on  the sequential $Z'$ and E6 models using
the charged lepton channels.  
The units used for mass limits are GeV$/$c$^2$ and for $\int {\cal L} \cdot dt$ are pb$^{-1}$.}
\label{tab:ZprimeTraditionLimits}
\begin{tabular}{llllll}
\hline\noalign{\smallskip}
Sequential $Z'$ &   $ee$  &  $\mu\mu$   &  $ee+\mu\mu$  & $\tau\tau$ & $\int {\cal L} \cdot dt$ \\
\noalign{\smallskip}\hline\noalign{\smallskip}
CDF                           & 750  & 735  & 815     &  394     &   200      \\
CDF with $\cos {\theta}$      & 845  &      &         &          &   448     \\
D\O \                            & 780  & 680  &         &          &  200-250 \\
\hline\noalign{\smallskip}
                              &      &      &         &          &            \\
\hline\noalign{\smallskip}
E6                           & $Z_l$        &  $Z_X$  &   $Z_\Psi$ &  $Z_\eta$    & Channel\\
\hline\noalign{\smallskip}
CDF                           & 615  & 675  & 690     &  720     &  $ee+\mu\mu$      \\
CDF with $\cos {\theta}$      & 625  & 720  & 690     &  715     &   $ee$    \\
D\O \                            & 575  & 640  & 650     &  680     &   $ee$   \\
\noalign{\smallskip}\hline
\end{tabular}
\end{table}

\subsection{Quark-Lepton Compositeness}
\label{sec:Compositeness}

\begin{figure}
\resizebox{0.5\textwidth}{!}{%
  \includegraphics{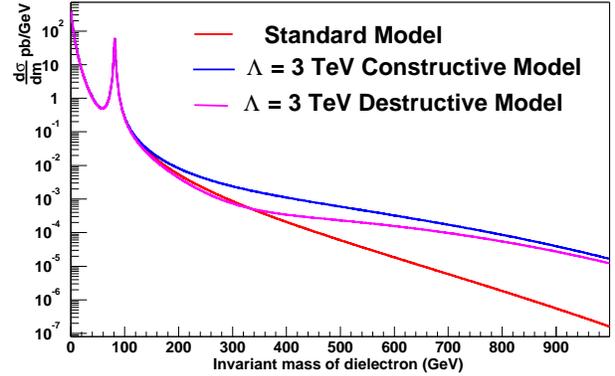}
}
\caption{$M_{ee}$ distributions for SM dielectron production and
for constructive and destructive interference due to contact
interactions.}
\label{fig:CompositeTheory}       
\end{figure}

Contact Interaction composite models introduce hypothetical
constituents of quarks and leptons called ``preons'' which
are bound together by a characteristic energy scale known
as the compositeness scale ($\Lambda$)\cite{bib:CompositeTheory}.
The differential cross-section can be written as in 
Equation~\ref{eq:Compositeness}.
\begin{equation}
\frac{d\sigma_T}{dM}=\frac{d\sigma_{SM}}{dM} + \frac{I}{\Lambda^2} + \frac{C}{\Lambda^4}
\label{eq:Compositeness}
\end{equation}
For energies accessible at the Tevatron,
the interference term (the second term) dominates
and quark-lepton compositeness
would be discovered as an excess in the tail of the dilepton distributions,
an example of which is shown in Figure~\ref{fig:CompositeTheory}.

\begin{figure}
\resizebox{0.5\textwidth}{!}{%
  \includegraphics{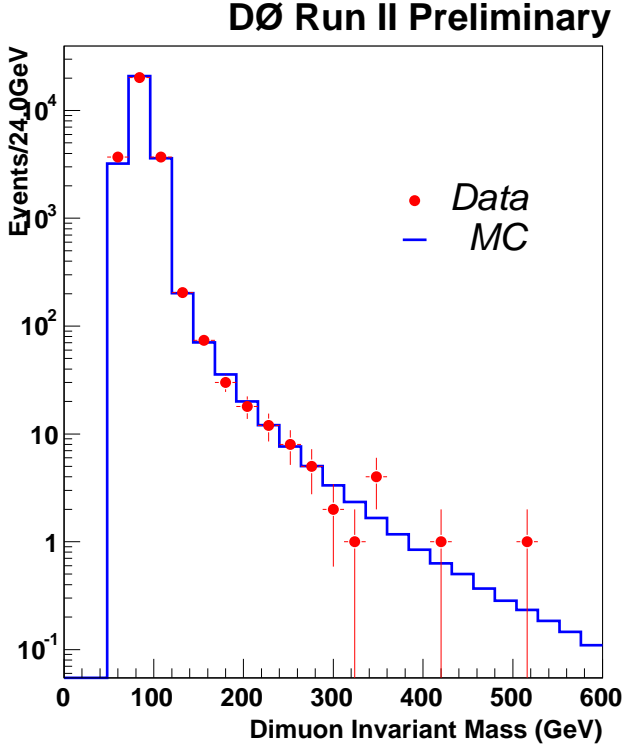}
}
\caption{$M_{\mu\mu}$ distribution for using $400\ $pb$^{-1}$ 
of data collected at D\O .}
\label{fig:CompositeMuon}       
\end{figure}

No evidence for signal is found in a dielectron 
search of $271\ $pb$^{-1}$ or in a dimuon search of $400\ $pb$^{-1}$ 
at D\O .  The dimuon results are shown in Figure~\ref{fig:CompositeMuon}.
Limits are set on $\Lambda$ for several models as shown in 
Table~\ref{tab:CompositeLimits}.

\begin{table}
\caption{Limits on the compositeness scale for several models.}
\label{tab:CompositeLimits}
\begin{tabular}{llllll}
\hline\noalign{\smallskip}
Model  &  $\Lambda-$&(TeV)   & & $\Lambda+$ &(TeV) \\
       &  $ee$ & $\mu\mu$ &  &  $ee$ & $\mu\mu$ \\ 
\noalign{\smallskip}\hline\noalign{\smallskip}
LL    & 6.2 & 6.9 &  & 3.6 & 4.2 \\
\hline
RR    & 5.8 & 6.7 &  & 3.8 & 4.2 \\
\hline
LR    & 4.8 & 5.1 &  & 4.5 & 5.3 \\
\hline
RL    & 5.0 & 5.2 &  & 4.3 & 5.3 \\
\hline
LL+RR & 7.9 & 9.0 &  & 4.1 & 5.0 \\
\hline
LR+RL & 6.0 & 6.1 &  & 5.0 & 6.4 \\
\hline
LL-LR & 6.4 & 7.7 &  & 4.8 & 4.9 \\
\hline
RL-RR & 4.7 & 7.4 &  & 6.8 & 5.1 \\
\hline
VV    & 9.1 & 9.8 &  & 4.9 & 6.9 \\
\hline
AA    & 7.8 & 5.5 &  & 5.7 & 5.5 \\
\noalign{\smallskip}\hline
\end{tabular}
\end{table}

\subsection{Extra Dimensions}
\label{sec:ed}

\subsubsection{Large Extra Dimensions}
\label{sec:led}

\begin{figure}
\resizebox{0.5\textwidth}{!}{%
  \includegraphics{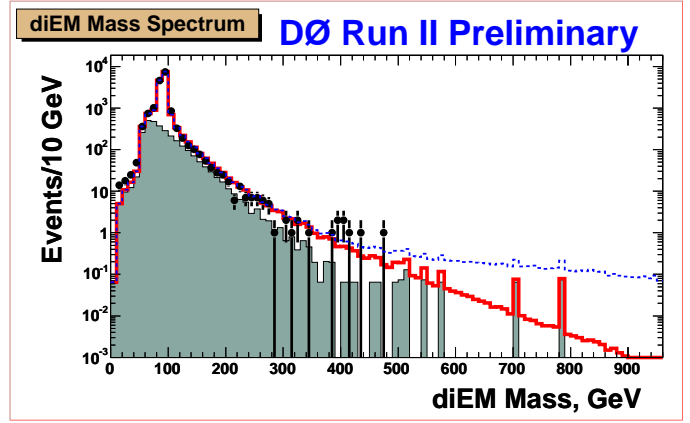}
}
\caption{Background prediction and observation of 
$M_{ee}$, $M_{\gamma\gamma}$  
distributions.  The dotted blue spectrum shows the
LED theoretical prediction for $\eta_G=0.6$.}
\label{fig:LEDmass}       
\end{figure}

\begin{figure}
\resizebox{0.5\textwidth}{!}{%
  \includegraphics{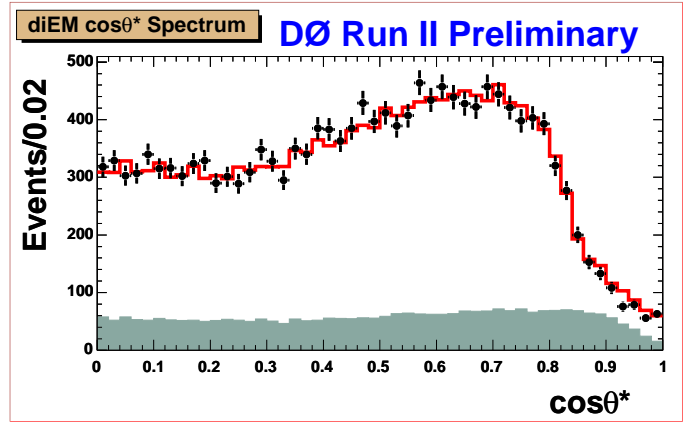}
}
\caption{$\cos{\theta^*}$ predicted and observed distributions for 
$ee$ and $\gamma\gamma$.}
\label{fig:LEDcostheta}       
\end{figure}

Large Extra Dimensions (LED) provide a non-SUSY alternative solution to the
``hierarchy'' problem in the SM and an explanation for the large 
difference between the electroweak and Planck scales
($M_{EW}<<M_{Pl}$).  The signature for LED
is dilepton or diphoton production.  The Large ED (ADD)
model\cite{bib:LedTheory} predicts an increase in
cross-section at high mass and depends on parameter
$\eta_G=F/M_s^4$ where $F$ is a model dependent dimensionless
parameter and $M_s$ is the UV cutoff, $M_s=M_{Pl(4 + n\ dim)}$.
An example $M_{ee}+M_{\gamma\gamma}$ distribution for $\eta_G=0.6$ 
is shown in Figure~\ref{fig:LEDmass} along with the
background prediction and observed data for $200\ $pb$^{-1}$
of dielectron and diphoton data at D\O .  
Figure~\ref{fig:LEDcostheta} shows no anomaly in the $ee$,
$\gamma\gamma$ $\cos{\theta^*}$ distribution.
By fitting $M_{ee}$, $M{\gamma\gamma}$,
and  $\cos{\theta^*}$, D\O \ extracts limits on $\eta_G$
at the 95\% C.L. such that $\eta_G^{95\%} < 0.292$ TeV$^{-4}$ 
for $\lambda>0$ and $\eta_G^{95\%} > -0.432$ TeV$^{-4}$ for
$\lambda<0$.

\subsubsection{Warped Extra Dimensions}
\label{WED}

\begin{figure}
\resizebox{0.5\textwidth}{!}{%
  \includegraphics{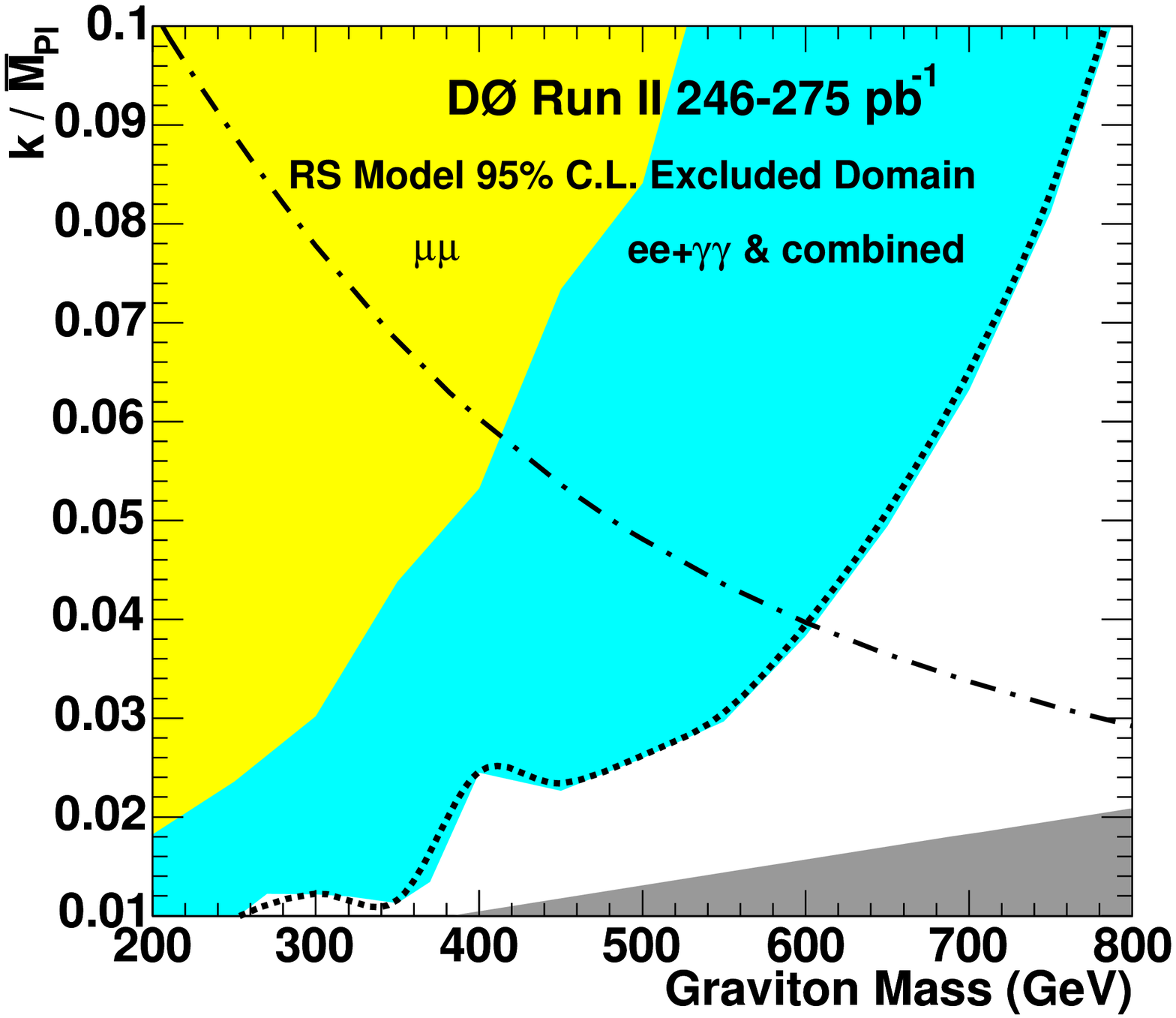}

  \includegraphics{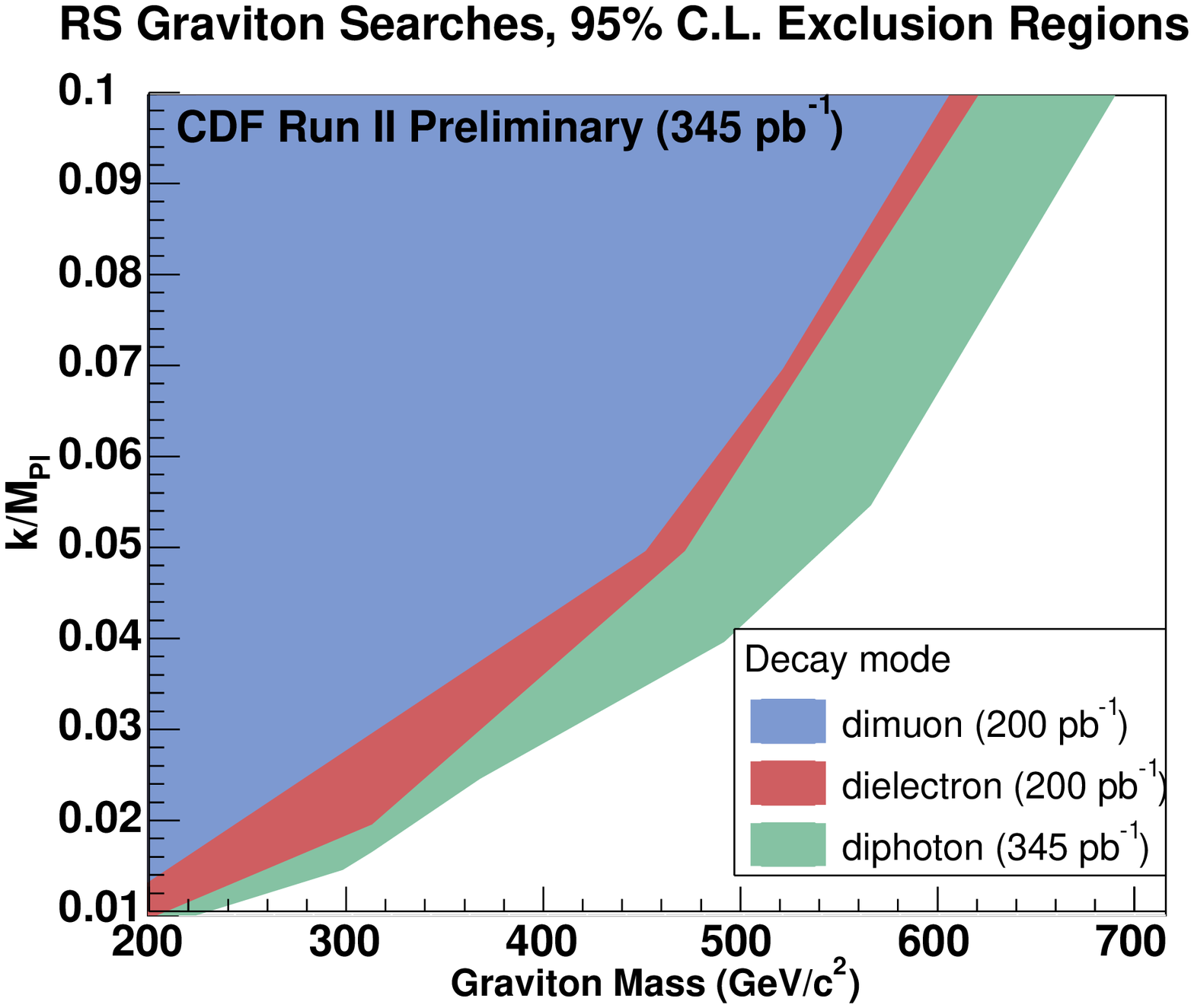}
}
\caption{Limits set on Randal Sundrum Graviton production
at D\O \ and CDF.}
\label{fig:RSGlimits}       
\end{figure}

The Warped Extra Dimension model predicts one extra
dimension that is highly curved and the production of 
Randall-Sundrum (RS) gravitons\cite{bib:RSGraviton}.  
The model depends on $k/M_{Pl}$, where $k$ is the
curvature scale.   CDF and D\O \ search for RS gravitons
by studying the $M_{ee}$, $M_{\mu\mu}$, and $M_{\gamma\gamma}$
distributions for a resonance which would depend on $k/M_{Pl}$.  
Two-dimensional exclusion regions in the $k/M_{Pl} -  M_{G}$ 
plane are established as shown in Figure~\ref{fig:RSGlimits}.

\section{Charged Heavy Vector Boson ($W'$)}
\label{sec:Wprime}

\begin{figure}
\resizebox{0.5\textwidth}{!}{%
  \includegraphics{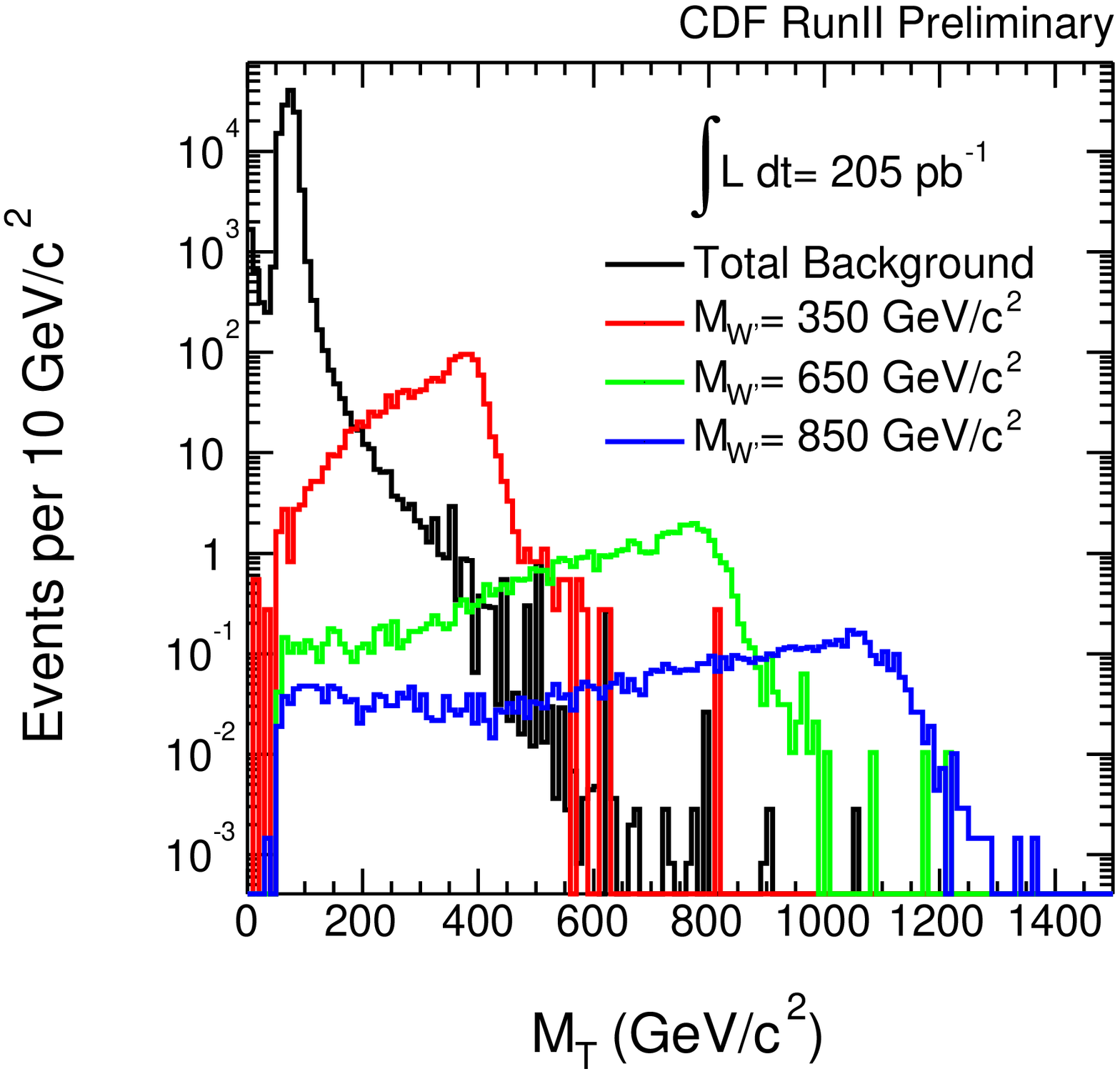}
  \includegraphics{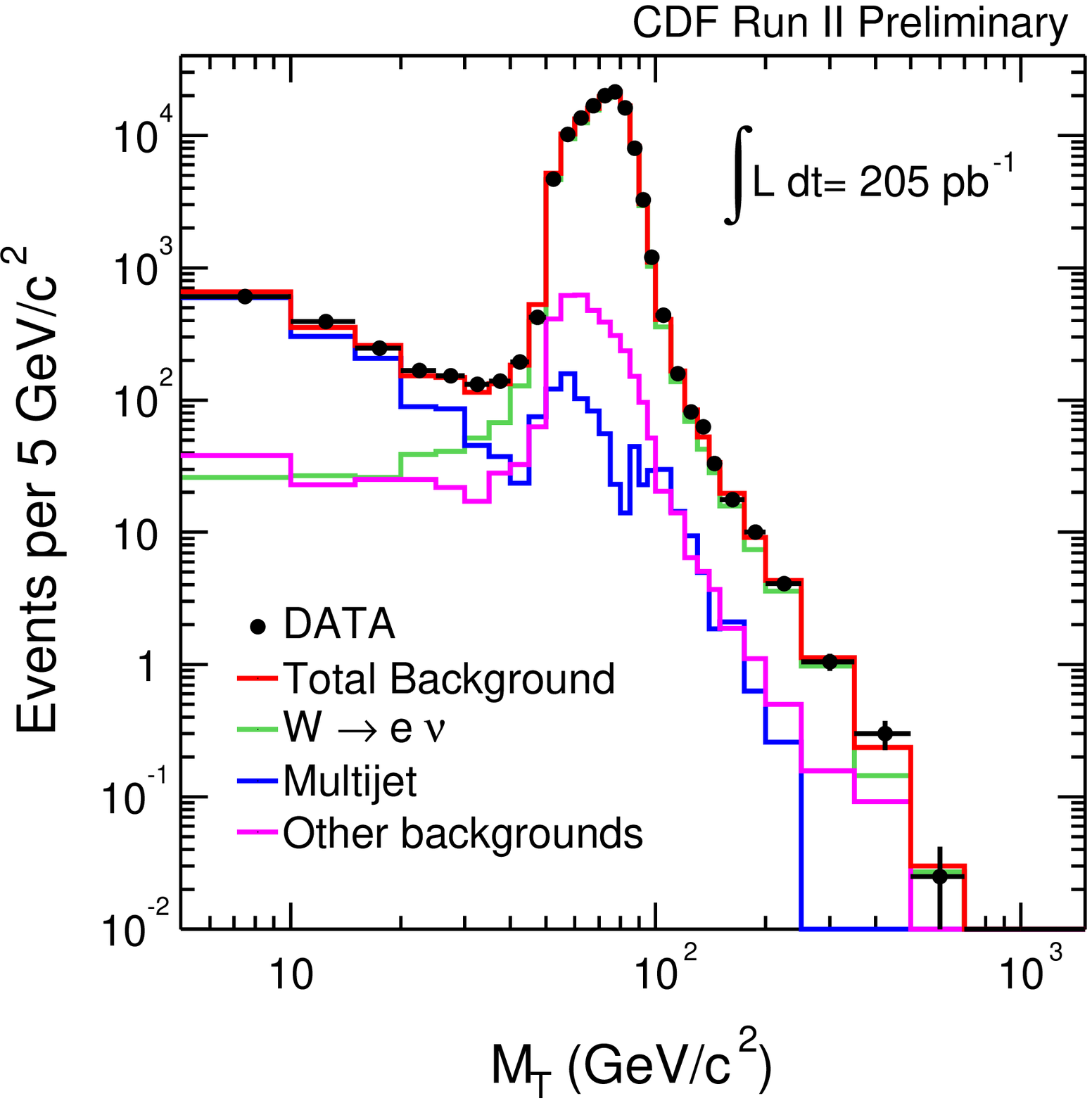}
}
\caption{The left plot has transverse mass distributions of 
the expected background overlaid with  
three $W'$ mass choices.  The right plot shows the transverse mass 
distributions of the irreducible 
SM $W\rightarrow e \nu$, multijet, and total background sources.
The data is plotted and agrees well with the expectation.}
\label{fig:wprime_mc}       
\end{figure}

The production of charged heavy vector bosons, referred to
as $W'$ particles, are predicted in theories based on
the extension of the gauge group\cite{bib:Wprime}.  The $W'$ is modeled
to decay to an electron and neutrino, where the neutrino
is assumed to be SM-like:  light and stable.
Thus, the final state signature in the detector
is a high $p_T$ electron with high missing $E_T$.
CDF performs a direct search for $W'$ production and
Figure~\ref{fig:wprime_mc} shows the background due
to SM $W\rightarrow e\nu$ production with the predicted 
transverse mass distributions
for $W'$ production at three different $W'$ masses.

Figure~\ref{fig:wprime_mc} shows the expected background
distributions and the observations in the data.  No 
$e\nu$ signal above the SM expectation is observed. 
However, the agreement between the data and the 
background prediction indicate good understanding of the
calorimeter energy at CDF and the detector missing energy.   

%
Having observed no signal above the SM expectation, the limit
at the 95\% C.L. is set on $W'$ production using a binned likelihood fitting
method.  
The CDF Run II
search excludes $W'$ masses less than 842 GeV$/$c$^2$. 
The CDF Run I limit was $M_{W'_{SM}} > 754$ GeV$/$c$^2$.
%
%
\section{Leptoquarks}
\label{sec:leptoquarks}

Many extensions of the SM assume additional symmetry between
lepton and quarks which requires the presence of a 
``new'' particle, a leptoquark (LQ)\cite{bib:LQ}.  Leptoquarks,
which could be scalar or vector particles, carry both
lepton and baryon numbers.  They are assumed to couple 
to quarks and leptons of the same generation; thus,
there are three generation of leptoquarks for which one
could search.

Leptoquarks would be pair produced at the Tevatron.  Their
decay is controlled by parameter $\beta$, where
$\beta = B.R.(LQ \rightarrow lq)$.
There are three final state signatures for LQ pair production 
at the Tevatron: two charged leptons 
and two jets ($lljj$); one charged, one neutral lepton and two jets 
($l \nu jj$); and two neutral leptons and two jets ($\nu\nu jj$).  
The experimental signal is a resonance in the lepton-jet
invariant mass spectrum.

\begin{figure}
\resizebox{0.5\textwidth}{!}{%
  \includegraphics{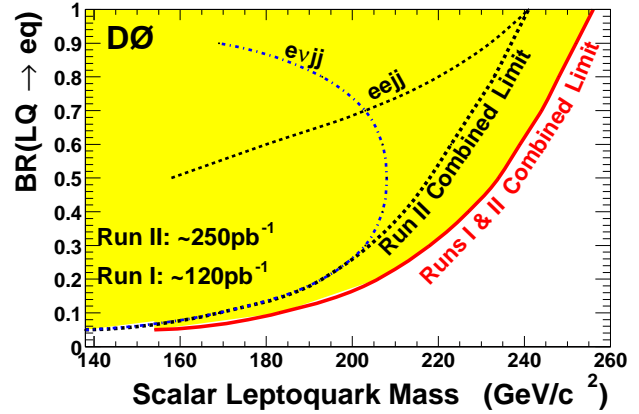}
}
\caption{Exclusion region established by D\O \ for first generation
leptoquarks.}
\label{fig:LQgen1}
\end{figure}

No evidence of LQ production is found at D\O \ or CDF.  
Figure~\ref{fig:LQgen1} shows the two dimensional exclusion
region established by D\O \ for the first generation with $eejj$ and $e\nu jj$
final state signature.  D\O \ combines 250 pb$^{-1}$ from Run II
with 120 pb$^{-1}$ of data from Run I to obtain the exclusion
region shown in Figure~\ref{fig:LQgen1}.  For the case
of $\beta=1$, D\O \ excludes first-generation leptoquarks with
masses less than 256 GeV$/$c$^2$.  CDF excludes masses less 
than 235 GeV$/$c$^2$ using 200 pb$^{-1}$ from Run II.

\begin{figure}
\resizebox{0.5\textwidth}{!}{%
  \includegraphics{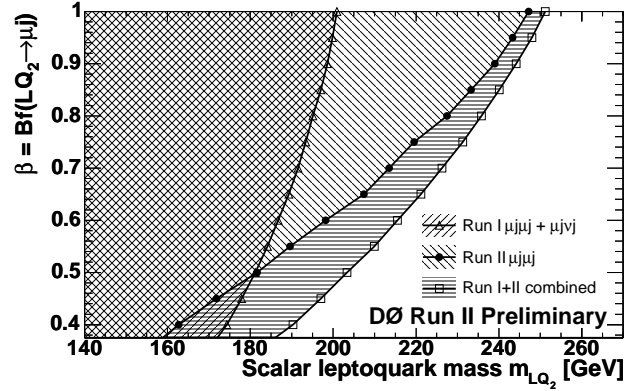}
}
\caption{Exclusion region established by D\O \ for second generation
leptoquarks.}
\label{fig:LQgen2}
\end{figure}

Figure~\ref{fig:LQgen2} shows the exclusion regions 
for generation two leptoquarks from D\O \ .
D\O \ searches for $\mu \mu j j$ and $\mu \nu j j$ production;
CDF searches for $\mu \mu j j$, $\mu \nu j j$, and $\nu \nu j j$ 
production.  
For $\beta = 1$, D\O \ Run I + II excludes LQ masses
less than 251 GeV$/$c$^2$ while CDF Run II excludes mass
less than 224 GeV$/$c$^2$.

CDF has performed a search for third generation 
LQ production using the $\tau\tau b b$ signature.
Leptoquark masses less than 129 GeV$/$c$^2$ are excluded
for $\beta=1$ using 200 pb$^{-1}$ of data. 

\section{Excited Electrons}
\label{sec:ExcitedElectrons}

The observation of excited states of leptons or quarks
would be a first indication that they are composite particles.
CDF searches for singly produced excited electrons ($e^*$)
in association with an oppositely charged electron, where
the $e^*$ decays to an electron and a photon.  Thus, the final
state signature is two electrons and a photon where the 
search signal is a resonance in the electron+photon 
invariant mass spectrum.

Two models are studied:  a Contact Interaction (CI) model\cite{bib:CIestar}
and a Gauge Mediated (GM) model\cite{bib:GMestar}.  The CI
model depends on the mass of the $e^*$ ($M_{e^*}$) and
the composite energy scale ($\Lambda$).  In the GM model,
an excited electron is produced via the decay of SM $\gamma^*/Z$.  This
model depends on $M_e^*$ and $f/\Lambda$, where $f$ is a phenomenological
coupling constant.

\begin{figure}
\resizebox{0.5\textwidth}{!}{%
  \includegraphics{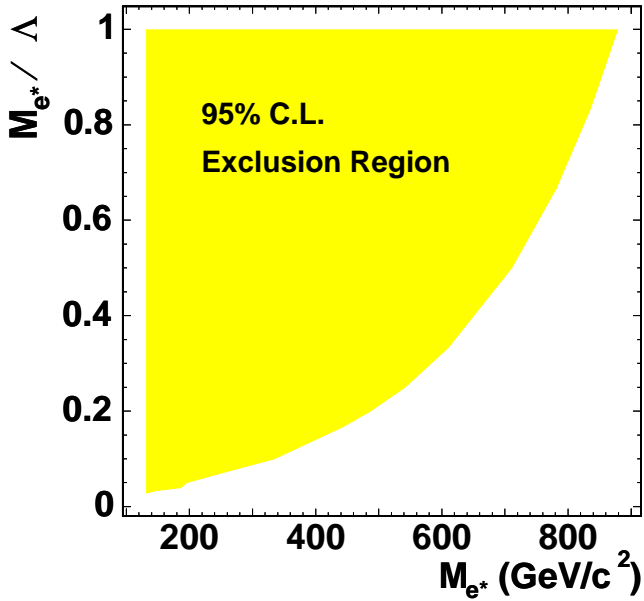}
}
\caption{Exclusion region at the 95\% C.L. established by CDF for $e^*$ production
via a Contact Interaction model.}
\label{fig:CIlimits}
\end{figure}

\begin{figure}
\resizebox{0.5\textwidth}{!}{%
  \includegraphics{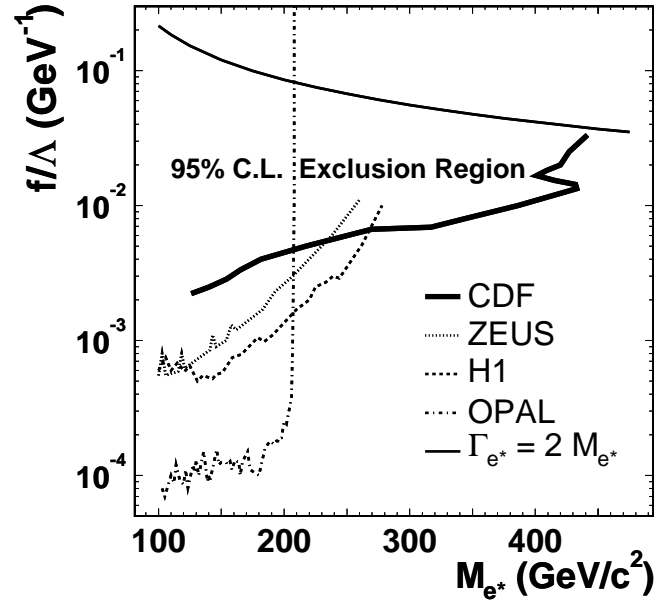}
}
\caption{Exclusion region at the 95\% C.L. established by CDF for $e^*$ production
via a Gauge Mediated model.}
\label{fig:GMlimits}
\end{figure}

In the first search for excited leptons at a hadron collider,
CDF found no excess of dielectron+photon events in 200 pb$^{-1}$ of data.  
Exclusion regions for each model are established.  Figure~\ref{fig:CIlimits}
shows the exclusion region at the 95\% C.L. in the
$M_{e^*}/\Lambda - M_{e^*}$ parameter space.  There are no previously 
published limits for $e^*$ production using the CI model.
For the GM model, it is conventional to plot
the 95\% C.L. exclusion region in the $f/\Lambda - M_{e^*}$
parameter space, as shown in Figure~\ref{fig:GMlimits}.  CDF
extends the previously published limits from 280 GeV$/$c$^2$ to 
$\approx$ 430 GeV$/$c$^2$.

\section{Summary}
Searches for physics beyond the Standard model using 200 pb$^{-1}$ to 
450 pb$^{-1}$ of data collected at CDF and D\O \ are presented.  Currently,
the experiments are actively persuing further exotic topics
and analyzing up to the full $1\ $fb$^{-1}$ of delevered luminosity.  
New and exciting results are coming out quickly.  Further information
regarding the analyses
presented in this paper and new results can be found at\cite{CDFresults}
and\cite{D0results}.


\begin{thebibliography}{}
%
%
\bibitem{bib:SeqZprimeModel}
J.Pati and A.Salam, Phys Rev Lett. \textbf{31}, 661 (1973);
R.N.Mohapatra, Phys. Rev. \textbf{D11}, 2558 (1975);
G.Sejanovic and R.N.Mohapatra, Phys. Rev. \textbf{D12}, 1502 (1975).
\bibitem{bib:E6ZprimeModel}
F.Del Aguila, M.Quiros, F.Zwirner, Nucl. Phys. \textbf{B287}, 419 (1987);
D.London and J.L.Rosner, Phys. Rev. \textbf{D34}, 5, 1530 (1986).
\bibitem{bib:carena}
M.Carena, A.Daleo, B.Dobrescu, T.Tait, Phys. Rev. \textbf{D70}, 093009 (2004).
\bibitem{bib:CompositeTheory}
E.Eichten, K.Lane and M.Peskin, Phys. Rev. Lett. \textbf{50}, 811 (1983);
E.Eichten, I.Hinchliffe, K.Late and C.Quigg, Ref. Mod. Phys. \textbf{56}, 579 (1984);
T.Lee, Phys.Rev. \textbf{D55}, 2591 (1997).
\bibitem{bib:LedTheory}
N.Arkani-Hamed, S.Dimopoulos, G.Dvali, Phys. Lett. \textbf{B429}, 263 (1998);
I.Antoniadis, N.Arkani-Hamed, S.Dimopoulos, G.Dvali, Phys. Lett. \textbf{B436}, 257 (1998);
N.Arkani-Hamed, S.Dimopoulos, G.Dvali, Phys. Rev. \textbf{D59}, 086004 (1999);
N.Arkani-Hamed, S.Dimopoulos, J. March-Russsell, SLAC-PUB-7949, e-Print Archive: hep-th/9809124.
\bibitem{bib:RSGraviton}
L.Randall and R.Sundrum, Phys. Rev. Lett. \textbf{83}, 3370 (1999).
\bibitem{bib:Wprime}
J.C.Pati and A.Salam, Phys. Rev. \textbf{D10}, 275 (1974);
R.N.Mohapatra and J.C.Pati, Phys. Rev.\textbf{D11}, 566 (1975);
G.Senjanovic and R.N.Mohapatra, Phys. Rev.\textbf{D12}, 1502 (1975);
\bibitem{bib:LQ}
M.Kramer, T.Plehn, M.Spira, P.M.Zerwas, Phys.Rev.Lett. \textbf{79}, 341 (1997).
\bibitem{bib:CIestar}
U.Baur, M.Spira, P.M.Zerwas, Phys. Rev. \textbf{D42},3 (1990).
\bibitem{bib:GMestar}
K.Hagiwara, D.Zeppenfeld, S.Komamiya, Z. Phys. \textbf{C29},115 (1985).
\bibitem{CDFresults}
http://www-cdf.fnal.gov/physics/exotic/exotic.html
\bibitem{D0results}
http://www-d0.fnal.gov/Run2Physics/WWW/results/np.htm
\end{thebibliography}
%

\end{document}